\documentclass[3p,twocolumn,times]{elsarticle}

\usepackage{ecrc}

\volume{00}

\firstpage{1}

\journalname{Nuclear Physics B Proceedings Supplement}

\runauth{S. R. Klein}

\jid{nuphbp}

\jnltitlelogo{Nuclear Science B - Proceedings Supplement}

\usepackage{amssymb}
\usepackage[figuresright]{rotating}

\begin{document}

\pdfmapfile{+txfonts.map}

\begin{frontmatter}



\dochead{}

\title{Radiodetection of Neutrinos}

\author{Spencer R. Klein}

\address{Nuclear Science Division, Lawrence Berkeley National Laboratory, Berkeley, CA, 94720 USA \\ and the University of California, Berkeley, CA, 94720 USA}

\begin{abstract}

Despite 100 years of effort, we still know very little about the origin of ultra-high energy cosmic rays.  The observation of neutrinos produced when cosmic-ray protons with energies above $4\times 10^{19}$ eV interact with the cosmic microwave background radiation, or in the neutrino sources, would tell us much about the origin and composition of these particles.   Over the past decade, many experiments have searched for radio waves emitted from the charged particle showers produced when EHE neutrinos interact with Antarctic or Greenland ice or the moon.  These experiments have not yet observed a neutrino signal.  Two groups are now proposing to instrument 100 km$^3$ of Antarctic ice with radio antennas, producing a detector large enough to observe a clear EHE neutrino signal in a few years of operation.  

\end{abstract}

\begin{keyword}

Neutrino \sep radiodetection \sep cosmic-ray 

\end{keyword}

\end{frontmatter}

% \linenumbers

\section{Introduction}
\label{Introxy}

Despite 100 years of effort, we do not know the source of high-energy cosmic rays.   For extremely high-energy particles (EHE, particles with energies above about $10^{17}$ eV), the mystery is even deeper.  We do not know of any likely  sources within our galaxy.  However, at the highest energies, above $4\times10^{19}$ eV (40 EeV), the range of cosmic-rays is very limited.  More energetic protons will interact with the $3^0$K cosmic microwave background radiation, and be excited to a $\Delta^+$ resonance, which will decay to $p\pi^0$ or $n\pi^+$, followed by $n\rightarrow pe^-\overline{\nu_e}$, as was first described by Greisen, Zatsepin and Kuzmin (GZK) \cite{GZK}.  The end result is a somewhat lower energy proton; this reaction recurs until the proton energy drops below 40 EeV.  For nuclei, a similar limitation is present, because the nuclei are photodissociated by the microwave photons.  These process limit the range of more energetic cosmic-rays to about 100~megaparsecs (Mpc).  Any sources must be relatively close.  

If they are protons, then the highest energy cosmic-rays should travel relatively straight lines bending a few degrees in 100~Mpc in the inter-stellar magnetic fields.  The Auger collaboration studied the arrival directions of EHE cosmic-rays.  They compared the arrival direction of cosmic-rays with energies above 60 EeV, with a list of active galactic nuclei within 75~Mpc of earth \cite{AugerAGN,Auger2}, and found a statistically significant correlation.  These correlations are only expected if cosmic-rays are mostly protons, because heavier nuclei will be bent in interstellar magnetic fields.  However, another Auger analysis indicates that, at energies above 10 EeV, cosmic-rays are mostly heavier nuclei \cite{Auger3}.

Clearly, an alternate probe of  EHE cosmic-rays is needed.  Neutrinos are that probe.  They are produced at any accelerator, by 'beam gas' interactions, when cosmic-rays undergoing acceleration collide with remnant gas and/or photons \cite{PT}.    Beyond that, there is a guaranteed source of EHE neutrinos. The $\pi^+$ produced by the GZK process decay, producing $\mu^+\nu_\mu$. Subsequently, the $\mu^+$ decays to $e^+ \overline\nu_e \nu_\mu$.  Over cosmic distances, these neutrinos oscillate, and arrive at Earth with a flavor mixture of roughly $\nu_e:\nu_\mu:\nu_\tau$ = $1:1:1$ (we neglect the difference between $\nu$ and $\overline\nu$).   Because these neutrinos interact weakly, and can travel cosmic-distances, Earthly  EHE neutrino detectors can observe neutrinos out to redshifts of 3-4 \cite{ESS}.  

The down side of the small cross-sections is that a very large detector is required to observe GZK neutrinos.  Optical detectors, like the 1 km$^3$ IceCube observatory \cite{RSI} are too small.  Analyses using the partially completed IceCube have set limits \cite{IC3EHE}, but even the completed detector is expected to observe less than one event/year.  A new approach is needed.

\section{Radiodetection of Neutrinos}

An alternative signature for EHE $\nu$ is the coherent radio Cherenkov emission from the electomagnetic and hadronic showers produced by their interactions.  The radio pulse comes about because of the Askaryan effect \cite{Askaryan}.  Any particle shower will have a net negative charge, due to Compton scattering of atomic electrons into the beam, and positron annihilation of atomic electrons.   In a medium, charged particles emit Cherenkov radiation.  When viewed at a wavelength that is large compared to the transverse size of the shower, the Cherkenov radiation adds coherently, and the radio signal scales as the neutrino energy squared.  For neutrino energies above $10^{17}$ eV, a substantial signal is present.  

The maximum frequency to maintain the coherence condition depends on the target material, but in ice, coherence begins to be lost above 1 GHz.   The radiation is emitted around the Cherenkov angle, about 55 degrees for radio waves in ice.  The angular width of the distribution depends on the longitudinal distribution of the shower particles, and on the frequency.  At very low frequencies, the radiation is nearly isotropic, while near the maximum coherence frequency, the radiation is heavily peaked at the Cherenkov angle.  

These properties of the radiation have been studied in a series of beam tests at SLAC \cite{beamtest}.   Beams of 25 GeV $e^-$ were directed into salt, sand and ice targets.  Total shower energies up to a few $10^{20}$ eV were studied.  However, it is important to remember that there are some differences between one $10^{19}$ eV electron and $10^{9}$ $10^{10}$-eV particles; at high energies, the Landau-Pomeranchuk-Migdal (LPM) effect reduces the electromagnetic cross-sections, and the electromagnetic shower length grows rapidly with energy.  As the showers become longer, the angular distribution of the radiation becomes more heavily peaked, and, eventually, at very high energies, useful coherence is lost.  To avoid this regime, some experiments have only considered radiation from the hadronic shower produced by the struck target.  This avoids some systematic uncertainty due to the LPM effect, but also throws away 80\% of the signal. 

Fortunately, at these energies, photonuclear and electronuclear interactions become important \cite{LPM}; they limit the growth in shower length.   At even higher energies, $\gamma$ coherently convert to $\rho^0$, which then interact hadronically \cite{coherence}. Detailed simulations are needed to account for these factors, and properly predict the radiation emission patterns; this would allow some of the neglected electromagnetic signal to be used.

\section{Radiodetection Experiments}

Over the past decade, many experiments have looked for radio signals from neutrino interactions in  a variety of targets: the moon, Antarctic ice, Greenland ice.  The FORTE satellite looked for 30-300 MHz radio waves as it passed over Greenland \cite{FORTE}.  Searches in large underground salt caverns have also been proposed \cite{SALSA}.  Most current effort is focused on the Moon and on Antarctic ice.  These two media are sensitive in different energy regions. 

\subsection{The Moon as a Target}

Because the moon is so far away, 
The moon is, of course, much larger than Antartica, so searches there can probe a larger target volume.  However, it is also much further away, and the inevitable signal spreading imposes an energy threshold above $10^{20}$ eV.  This is beyond the bulk of the GZK spectrum, so these experiments have largely been used to put limits on exotic, top-down models. 

Since the pioneering Parkes \cite{Parkes} and \cite{glue} GLUE experiments searched for microwave radiation from the Moon using large radio telescopes, many other radio astronomy facilities have been used to search for neutrino interactions.  Different experiments have searched in different frequency bands; there are advantages to both high and low frequency searches. 

One of the most sensitive low-frequency results came from the NuMoon experiment, which used eleven Westerbork 25 m dish antennas to study the 113-175 MHz frequency range.  At low frequencies, the radio emission is largely isotropic, so, the sensitive region covers the entire near-side surface of the moon.  Low-frequency experiments also probe deeper in the moon, since the radio attenuation length in the lunar regolith decreases with frequency; very roughly, the attenuation length is about 9~m/f(GHz).  NuMoon observed no events in 47~hours of observation, and set flux limits for neutrino energies above about $3\times10^{22}$ eV \cite{NuMoon}.  Observations are being continued by the low frequency array for radio astronomy (LOFAR), which is beginning to take data with 36 stations spread over Northwest Europe.  Each station comprises 48 antennas covering 120-240 MHz, plus 96~antennas which cover 10-80 MHz.  In the longer term, the proposed square kilometer array will continue this program; with 1 km$^2$ of collecting area, the threshold could be pushed down to near $10^{20}$ eV. 

Other experiments have searched for higher frequency emission.  For coherent emission, the radio power increases linearly with the frequency, so the signal is stronger.  However, the radiation is peaked near the Cherenkov angle, so high-frequency searches are sensitive only for a limited geometry, with the neutrino pointing in the correct direction, and interacting near the limb of the Moon, so that the radiation reaches the Earth.  They are thus sensitive to lower neutrino energies, but may have lower flux limits.

The Lunaska experiment at the Australia Telescope Compact Array (ATCA) took 6 nights of data with 6 22-m diameter dishes \cite{Lunaska}.  They studied the region between 1.2 and 1.8 GHz.  Because of the wide bandwidth, the experiment needed a filter to remove the signal dispersion in the Earth's atmosphere.    The Resun experiment took 45 hours of data with the very large array, in 50 MHz bandwidth around 1.45 GHz \cite{Resun}.  A more advanced follow-on experiment is planned.

\subsection{Antarctic Experiments: RICE and ANITA}

\begin{figure} [tb]
\centering
\includegraphics[totalheight=0.35\textheight]{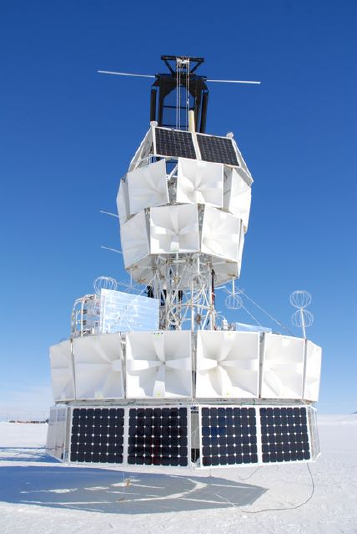}
\caption{A Photo of the ANITA Balloon experiment, awaiting launch.  The white squares are the horn antennas, and the darker rectangles are solar panels for power.  From Ref. \cite{ANITA}.}
\label{fig:ANITA}
\end{figure}

In contrast, Antarctic experiments can have a much lower threshold, down to $10^{17}$ eV.   The first Antarctic experiment was the Radio Ice Cherenkov Array (RICE), which deployed antennas in the holes drilled by the AMANDA experiment, at the South Pole \cite{Rice}.  It set limits on neutrinos with energies above 50 PeV, and  pioneered many of the techniques used by later in-ice experiments.

More recently, the ANITA balloon experiment shown in Fig. \ref{fig:ANITA}, made two flights around Antarctica, at an altitude of about 35,000 m, looking for signals from neutrino interactions in the ice below.  The detector looked out to the horizon, up to 650 km away, so the experiment had a very large sensitive volume.  The collaboration has analyzed data from a 35 day flight in 2006/7, and a 31~day flight in 2008.   The balloon carried 40 horn antennas (32 in the first flight), each read out by low-noise amplifiers and 2.6 GS/s waveform digitizers.  Each horn had two readout channels, for horizontal and vertical polarization.  The polarization sensitivity is important, since real signals are expected to be vertically polarized. 

By comparing the signal arrival times at different antennas, ANITA had a pointing accuracy of about $0.2^\circ$ - $0.4^\circ$ in elevation and $0.5^\circ$  - $1.1^\circ$ in azimuth, depending on the signal size.  The detector was calibrated using buried transmitters which measured the signal propagation through the ice, firn, and firn-air interface.  

During the 2nd flight, ANITA collected a total of 26.7~million triggers.  Readout was triggered when 4 antennas (two in the upper rings, and two in the lower) registered signals well above the ambient noise level.   During the second flight, the noise levels were automatically adjusted, and antennas pointing toward known noise source (e.g. Amudsen-Scott South Pole Station) could be masked from the trigger.  Out of the 26.7 million triggers, about 320,000 could be reconstructed as a point source, in target ice, within the fiducial angle.  At this stage, the largest remaining background seen by ANITA was anthropogenic (man-made) signals, from the South Pole station and other inhabited sites.  Other significant backgrounds were payload noise, thermal noise, and misreconstructed events.  The collaboration applied a series of cuts to remove this background.  After cuts to reject this background, five events remained from the 2nd flight dataset.  Three of these were horizontally polarized, and were likely radio signals from cosmic-ray air showers  \cite{ANITA,ANITAcr}.  The other two published events were consistent with the expected signal; this figure should be compared with the estimated background of $1.0\pm 0.4$ events; it was later determined that one of the signal events was actually an artificial pulser event \cite{ANITA}.    From this, the ANITA collaboration has set upper limits which begin to constrain interesting' GZK models. 

\begin{figure} [tb]
\centering
\includegraphics[totalheight=0.35\textheight]{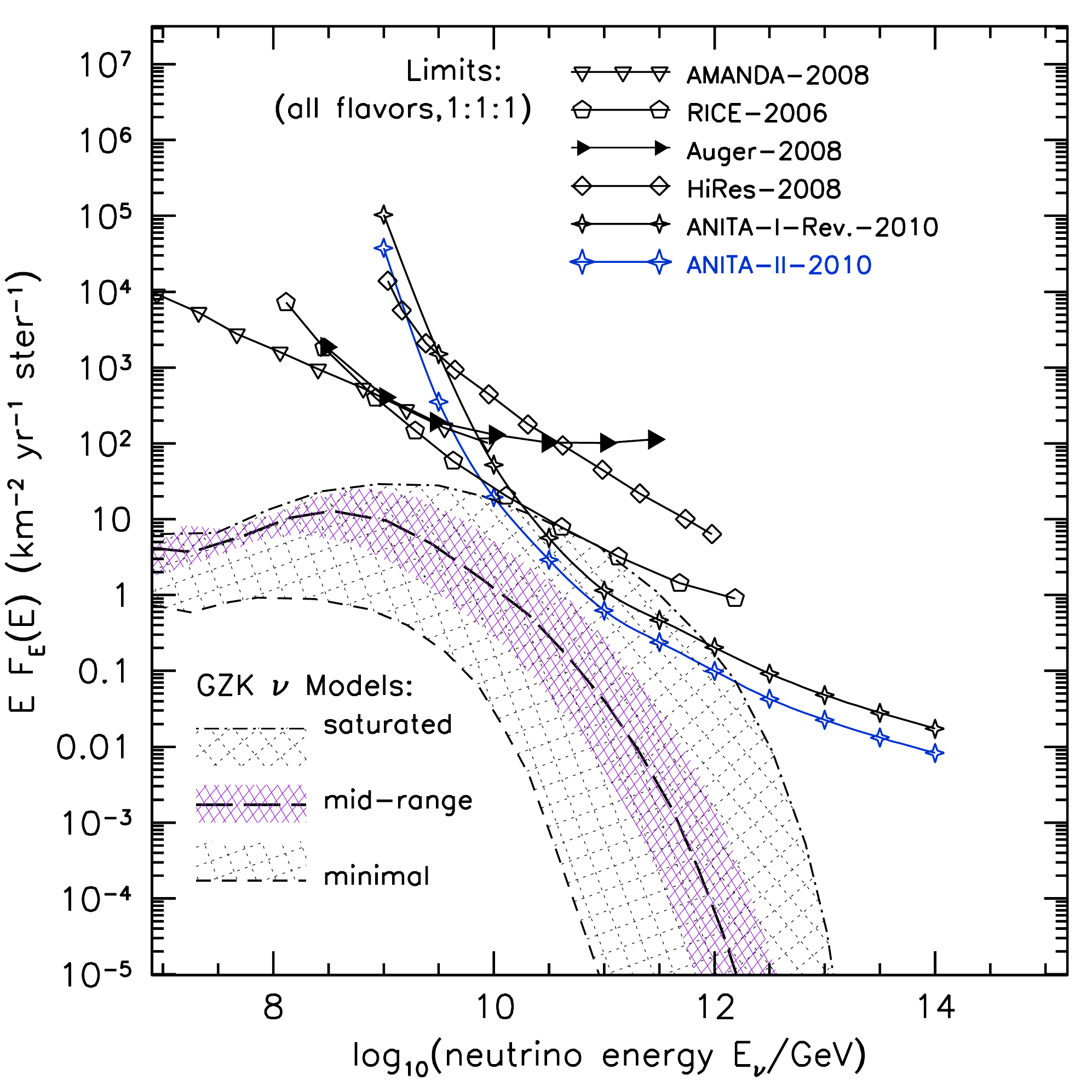}
\caption{The corrected EHE $\nu$ flux limits set by the ANITA experiment, compared to some other recent measurements.  The shaded band shows the range of $\nu$ fluxes expected in GZK models.   From Ref. \cite{ANITA}.}
\label{fig:ANITAlimit}
\end{figure}

The final limits set by ANITA are shown in Fig. \ref{fig:ANITAlimit}, along with results from RICE and some other experiments.   The collaboration has been approved for a third flight in 2013-2014.

\section{Future In-Ice Detectors}

A key goal of future detectors is to reach a lower energy threshold, around $10^{17}$ eV, so as to be able to be sensitive to the bulk of the GZK $\nu$ spectrum.  This will remove any systematic uncertainties due to the spectrum shape, and, in the longer run, allow for a spectral measurement.    To reach an energy threshold around $10^{17}$ eV, it is necessary to locate antennas in the active volume.  These experiments also target a volume of order 100 km$^3$, enough to observe ~ 100 GZK $\nu$ in 3-5 years. Two groups are pursuing this approach; both experiments are at the prototype stage.

\begin{figure*} [tb]
\centering
\includegraphics[totalheight=0.25\textheight]{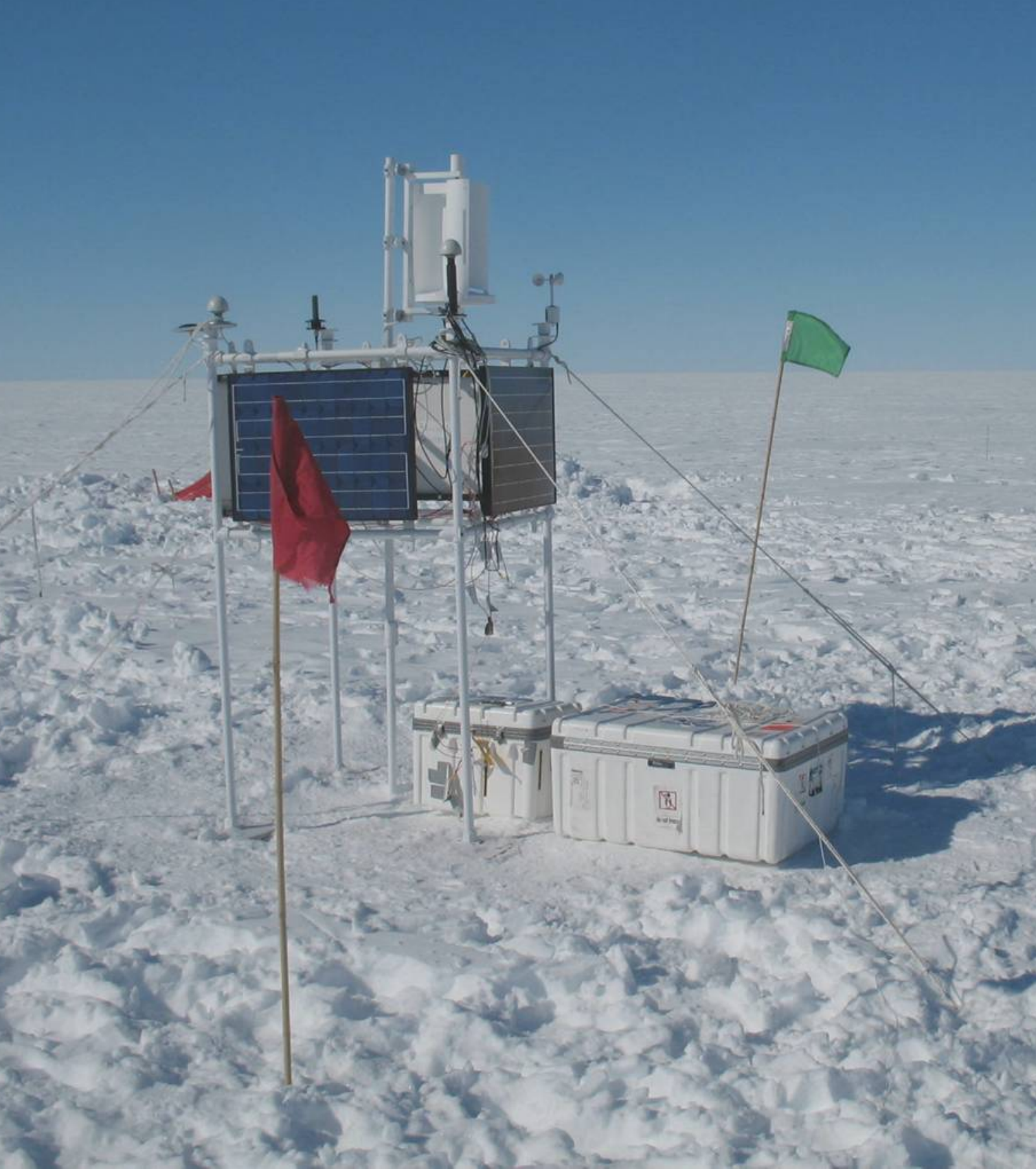}
\includegraphics[totalheight=0.25\textheight]{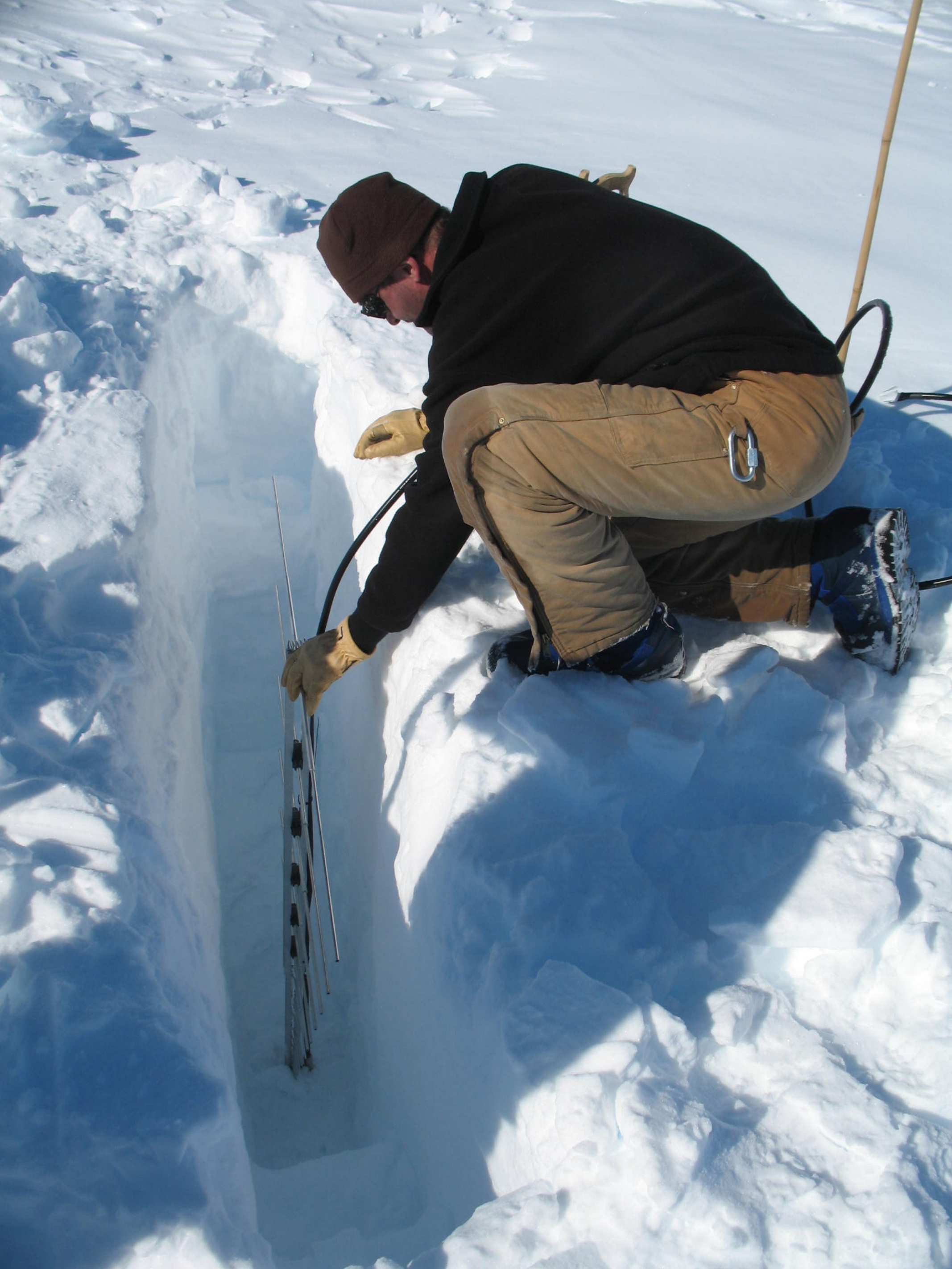}
\includegraphics[totalheight=0.25\textheight]{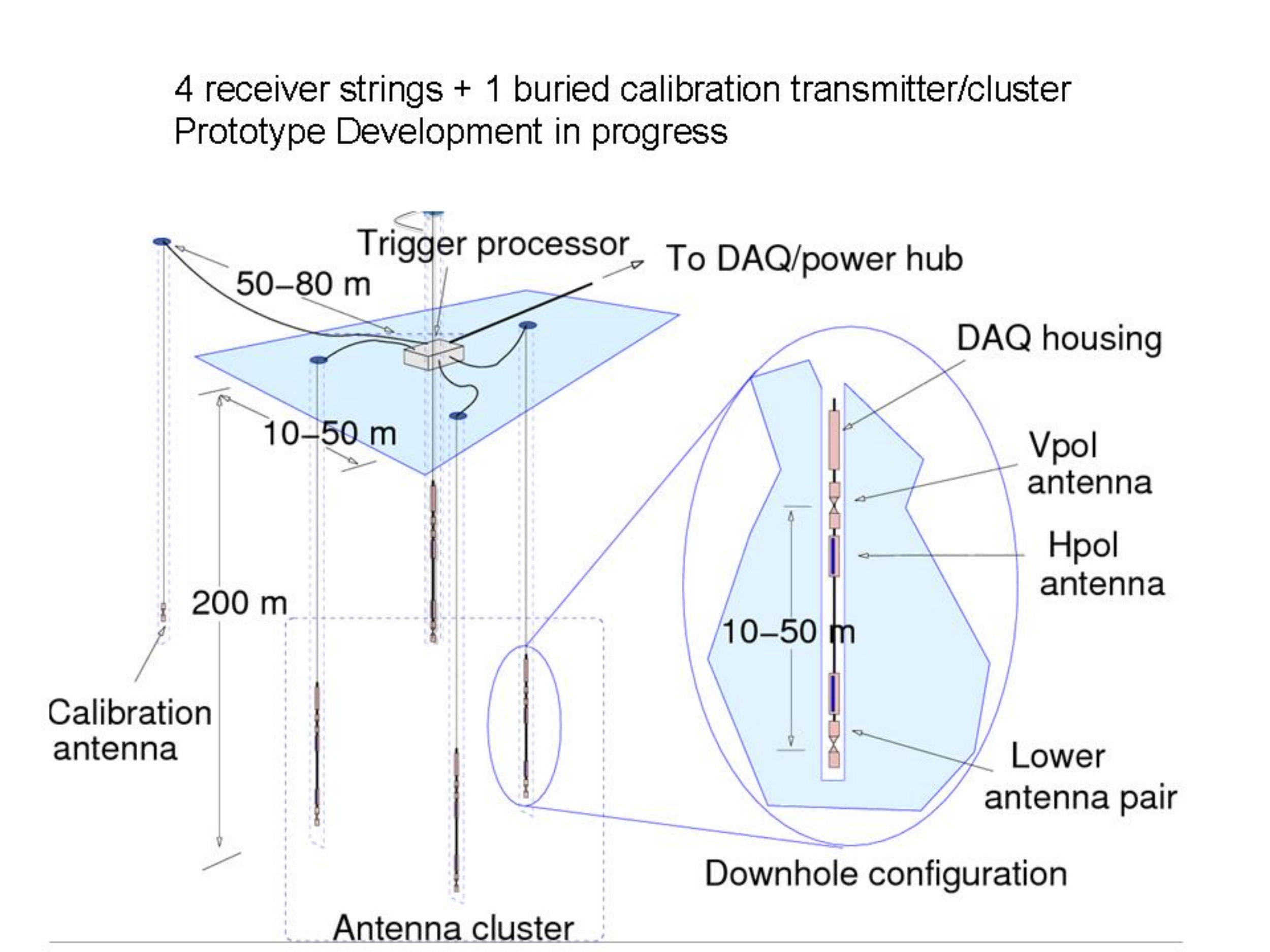}
\caption{(left) The ARIANNA prototype station, during deployment in Moore's Bay.  The white boxes contain the electronics and battery box; these components were later buried. (center).  One of the ARIANNA antennas being buried (right) Schematic layout of an ARA detector cluster.}
\label{InIce}
\end{figure*}

Askaryan Radio Array (ARA) is proposing to deploy clusters of radio antennas in 200 m deep holes at the South pole \cite{ARA}.  As Fig. \ref{InIce}(right) shows, the clusters would be located on a 1-1.5 km triangular grid.  Each hole would contain several antennas, allowing for local up-down discrimination.  The South Pole ice is cold, so the radio attenuation lengths are more than 500 m.  The group plans to instrument the antennas with conventional waveform digitizers, although possible variations are under discussion, whereby the number of antennas would be increased, but with each antenna read out by simpler time-over-threshold electronics. The collaboration has deployed a series of prototypes receivers and radio transmitters in holes drilled for IceCube.  

The Antarctic Ross Ice Shelf Antenna Neutrino Array (ARIANNA) is proposing to deploy a series of stations in a grid in Moore's Bay on the Ross Ice Shelf.   There, 572 m of ice sits on top of the Ross Sea.  One key advantage of this site is that the ice-water interface reflects radio waves.  This is a key advantage, in that it gives the array sensitivity to downward going neutrinos, as well as improved sensitivity to more horizontal tracks.  Also, the firn is shallower, with a more rapid transition to solid ice \cite{firnpaper}.  These factors allow antennas to be deployed just below the surface.

Each ARIANNA station will consist of 8 log-periodic antennas buried, pointing vertically downward, in the ice. The antennas cover 105 to 1300 MHz in air, and go down to slightly lower frequencies when embedded in the ice.  The antennas will be read out by waveform digitizer systems sampling at about 2 GS/s.  They are triggered whenever 2 out of 4 antennas record a supra-threshold signal.  The trigger in the prototype system divided the input signals up into two frequency bands, with independent thresholds; this may not be cost effective in the full array.

ARIANNA established field camps at the site in 2007 and 2009.  In 2007, the radio attenuation length was measured by reflecting radio waves off the ice-water interface.  The attenuation length is in the 300-500 m range, depending on frequency \cite{ARIANNAradio}.  in 2009, a prototype station was deployed.  It is shown in Fig. \ref{InIce} (left), and one of the antennas is shown in Fig. \ref{InIce}.  it collected data until the sun set in March; after the wireless internet was removed, no anthropogenic triggers were observed. 

Some of the biggest problems for these Antarctic radiodetection are common to both experiments.  The biggest one may be power during the winter.  During the Antarctic summer, solar panels provide adequate power.  However, batteries do not work well at Antarctic temperatures, and cannot sustain experiments through the 6-month long winter.  Both groups are exploring the possibility of wind generation at their sites.   Both experiments do not plan to run cables between the stations, so wireless communications is needed, in a form that does not interfere with signal detection.  This should be, in principle possible by using higher frequency (5 GHz?),  wireless communication  but it needs to be carefully checked.

Both experiments have plans for prototype arrays, to be installed within the next 2-4 years.  After that, presumably, one approach will be selected, and a large array built.

\section{Conclusions}

The observation of GZK neutrinos would finally give a definitive answer about the composition of EHE cosmic rays, and, at the same time, give us directional information about their probable sources.  However, because the EHE neutrino flux and cross-sections are small, they have not yet been observed. 
Two new experiments have been proposed to search for these neutrinos.  ARIANNA and ARA  will have active volumes of order 100 km$^3$.  If EHE cosmic rays are mostly protons, this is big enough to observe of order 100 neutrinos in 3-5 years of operation. 

This work was funded in part by the U.S. National Science Foundation under grant numbers 0653266 and 0969661 and the Department of Energy under contract number DE-AC-76-00098.

\section*{References}

\bibliographystyle{elsarticle-num}
\bibliography{<your-bib-database>}

\end{document}